\documentclass[]{article}

\usepackage[]{fullpage}
\usepackage{lineno,hyperref}
\usepackage[utf8]{inputenc}
\usepackage{hyphenat}
\modulolinenumbers[5]

\usepackage{amsmath}
\usepackage{graphicx,authblk}

\usepackage{fullpage}

\newcommand{\ie}{\emph{i.e.~}}

\begin{document}

\title{Modelling risk-taking behaviour of avalanche accident victims\\
\small \emph{(Research Master thesis - M2 Cognitive Sciences, PHELMA, Univ. Grenoble-Alpes)}}

\author[1]{Robin Couret}
\author[1,2]{Carole Adam}
\author[1,3]{Martial Mermillod}

\affil[1]{Univ. Grenoble-Alpes, Grenoble, France}
\affil[2]{Grenoble Informatics Laboratory, CNRS UMR 5217}
\affil[3]{Laboratoire de Psychologie et Neurocognition,
CNRS UMR 5105}

\date{August 2020}
\maketitle

\abstract{
Each year, over 15000 requests for mountain rescue are counted in France. Avalanche accidents represent 39\% of reports, and are therefore our focus in this study. Modelling the behaviour of mountain accident victims is useful to develop more accurate rescue and prevention tools. Concretely, we observe the interference of two heuristics (availability and familiarity) in decision making when choosing an itinerary in backcountry skiing. We developed a serious game to evaluate their effect on the probability of engaging in a risky itinerary, while varying situational and environmental criteria in each participant (N = 278). The availability heuristic is operationalized by three situations, an avalanche accident video, a backcountry skiing video and a neutral context. The familiarity heuristic is operationalized by two criteria, strong and weak familiarity with the place. Results demonstrate the effects of both heuristics. Measurements through our serious game are discussed in the perspective of developing an interactive prevention tool for mitigating the negative effects of heuristics.
}

\section{Introduction}

Each year, over 15000 requests for mountain rescue are counted in France \footnote{National Geographic and Forester Information Institute. CHOUCAS project presentation (2017). \url{http://choucas.ign.fr/\#presentation} (accessed 11 June 2020)} after various incidents or accidents.
An analysis of incident and accident reports filed on the SERAC database\footnote{SERAC database of incident and accident reports: \url{https://www.camptocamp.org/articles/697210/en/serac-database-of-incident-and-accident-reports}} shows that the majority (39 \%) of the reports refer to avalanche-type events
\cite{soule2017incidents}. In addition, the National Association for the Study of Snow and Avalanches \footnote{ANENA, accidents report: \url{http://www.anena.org/5041-bilan-des-accidents.html}} identified in France, over the winter seasons 2017-2018 and 2018-2019, 50 deaths and 85 injuries related to avalanche accidents. 
Modelling the behaviour of mountain accident victims is useful to develop more accurate rescue and prevention tools. Since avalanche-induced accidents are strongly represented in mountain accident reports, we have selected them as our study case for modelling human risk-taking behaviour leading to mountain accidents.

The location and rescue of mountain accident victims is based on a manual search for geographical clues, as well as on the know-how of the rescuers. This localisation task is complicated by the inaccuracies in the description by the victim of the route taken, which can be distorted by the affects and stress linked to the situation. The CHOUCAS project \footnote{CHOUCAS project: \url{http://choucas.ign.fr/}} aims to provide rescuers with methods and tools that will make it possible to improve the location of victims in the mountains. This study is part of this project, based on the assumption that understanding the decision making of accident victims can help locating them.
Besides, before an accident event, mountain activity practitioners make various decisions related to route choice. This decision-making may be associated with risks, and subject to errors in cognitive processing. To promote the least risky choices for mountain practitioners, a preventive approach can also be put in place, to make them aware of their possible errors in judging risks.

We approach risk taking through the perception of risk theory, which is a component of decision making under uncertainty \cite{slovic2004risk}. The perception of risk and decision making can be quantified by the subjective assessment of probability. In addition, \cite{tversky1974judgment} highlighted errors of judgment in humans which are produced by cognitive shortcuts called heuristics. Finally, in this study we therefore seek to understand the influence of heuristics on the choice of a ski route.

In order to study decision-making in mountain activities, we have created a dedicated experimental software to test the effects of heuristics on route selection. Our software is a web-based a serious game, in order to promote participant engagement \cite{lumsden2016gamification}. In addition, this video game approach provides a basis for playful prevention, the objective of which would be to reduce the effect of heuristics in off-piste skiing. All of the variables identified during the software development process may also be useful in the design of a computational model for simulating and understanding skiers' decisions.

\section{Theoretical background and hypotheses}

In order to improve the location and prevention of accidents in the mountains, we want to understand the reasoning behind the engagement in a risk situation. We have selected the avalanche accident situation as a case study for mountain accidents. On the one hand, we believe that this type of accident is often conditioned by the choices of the practitioner. And on the other hand we reduce the problem to the decision making that led to the accident.

In order to position the decision-making process in relation to an avalanche accident situation, we propose to combine the theoretical aspects of decision-making with the singular situation in which it takes place.
First, we approach decision making as the result of information processing which can be automatic or controlled. We then study how decision making is associated with risk taking. This results in the use of this theoretical approach to address the specific problem of mountain accidents.
Second, we discuss errors in judgment and their consequences on decision making. Subsequently, the focus is on certain specific cognitive processes that take place during off-piste skiing. Ultimately, this leads to postulating a set of hypotheses.

\subsection{Decision making: from the general approach to the case of off-piste skiing}

To make a decision is to choose between different options. It is a form of reasoning that applies in the activity of an individual and that depends on objectives and constraints relating to the situation \cite{meunier2016raisonnement}. These constraints result from the interaction between the environment and the individual. From the point of view of information theory, it is considered that information from the situation is processed by the body. Sensory information (\ie visual, auditory, etc.) is captured and then processed by the cognitive apparatus. Finally, information processing results in decision making and the selection of a response to execute \cite{wickens1991processing}. Decision making is therefore the product of cognitive processes, they can be automatic or controlled. This dichotomy of cognitive mechanisms of information processing refers to the dual process models \cite{kahneman2002representativeness}. In this theoretical framework, heuristic processing (or System 1) groups together automatic, unconscious and rapid processes which are inexpensive in cognitive resources but more prone to errors. In contrast, systematic processing (or System 2) groups together controlled, conscious and slow processes. The latter are more expensive in cognitive resources but they allow to perform high level tasks such as anticipation or planning \cite{evans2008dual}.

One of the components of perspective theory \cite{tversky1979prospect} is to establish that a heuristic treatment can result in a probability associated with a choice. In the case of a complex situation, the cognitive system has a limited probabilistic representation, so it favors heuristic processing which simplifies the perception of the situation \cite{tversky1983extensional}. This type of situation refers to decision making under uncertainty where knowledge of the consequences of a choice is not precise or non-existent. This is the case in a risk situation. While a risk can be defined as an event that can cause harm to humans \cite{leplat2006risque}, risk perception relates to the estimation of the probability of occurrence of an adverse event. It is then a question of the subjective assessment of a risk by an individual \cite{slovic1987perception}. Finally, this intuitive estimation of the situation is a process carried out by heuristic processing.

We take this general approach to decision making to model the risk taking of mountain accident victims. We reduce the problem to making a decision about which direction to take on a mountain journey. The situation is then viewed as a system with several route options. The choice of an itinerary is then associated with a probability of engagement as well as a risk. We then estimate that the probability of engagement is correlated with the perceived probability of occurrence of a risky event. We therefore wish to study the effect of heuristic processing on the probability of engagement in order to determine the errors in judgment leading to taking a risk.

\subsection{Heuristics}

We have explained above how heuristic processing can cause errors in judgment. In addition, to be more accurate, judgment heuristics can lead to a distortion of information known as cognitive bias. These phenomena were studied by Tversky and Kahneman in their work on judgment under uncertainty \cite{tversky1974judgment}. They then identify different heuristics that influence the evaluation of probabilities.

\subsubsection{Availability heuristics}

\begin{sloppypar}
One of the heuristics identified refers to situations where individuals assess the probability of an event. It is shown that the evaluation process is based on the ease with which events can be evoked. So since an event is more available, it is rated as more frequent, this is the availability heuristic \cite{tversky1973availability}. Thus the instances of an available event class are perceived as more numerous than the instances of an unavailable event class, for two classes of equal frequency. For example, the impact on the subjective perception of the probability of an accident seen or experienced is greater than the impact of reading information about this accident. The memory works in an associative mode which is reinforced by repetition. According to Tversky and Kahneman, the availability heuristic exploits the inverse form of this law, that is, it uses the strength of association as the basis for the judgment of frequency. This is why it is considered that the availability is higher when an event is more frequent. However, availability also depends on various factors unrelated to the actual frequency. Therefore Tversky and Kahneman believe that in real life one of the strongest demonstrations of availability heuristics is the impact of an unlikely accident.
\end{sloppypar}

More recent work associates the heuristic of availability with the heuristic of affect. Pachur compared the effect of these two heuristics on risk perception \cite{pachur2012people}. To do this, the author associates three indicators with risk:
\begin{itemize}
    \item the perceived frequency of a risk,
    \item the value of life (VSL) which refers to the financial cost necessary to reduce the number of deaths of a risk class,
    \item the perceived risk, introduced by Slovic \cite{slovic2000perception}.
\end{itemize}
The limitation of this experimental paradigm is that it is not intended for an ecological risk assessment task related to a particular situation. Rather, it makes it possible to compare the differences in perception between risks of different kinds. Thus in our study, for simplicity, we assume that the heuristic of availability and the heuristic of affect have a joint effect.

The availability heuristic is identified in cognitive processing associated with off-piste skiing. In this way the effect of the availability of an avalanche accident experience on the perception of risk was tested by Gletty \cite{gletty2017comprendre}. The results suggest that an off-piste ski practitioner with experience of an avalanche accident takes more account of avalanche risk information when deciding to go off-piste.

From the various works on availability heuristics and risk perception and wishing to reproduce results specific to the effect of the availability of an avalanche accident experience on risk perception, we postulate the following hypothesis:
\begin{quote}
    \textbf{Hypothesis 1:} In a decision-making context relating to the practice of off-piste skiing, the availability of an avalanche accident event for an individual decreases his probability of taking a route to avalanche risk.
\end{quote}

\subsubsection{Heuristics and information distortion}

As we have discussed, on the one hand heuristics can lead to distortion of information and on the other hand the different sensory sensors of the body select information from the environment. Therefore, in order to study the effect of heuristics, we believe that it is necessary to identify the information perceived from the environment.

In her work, \cite{gletty2017comprendre} identified a set of criteria that could be associated with the environment of an off-piste ski descent. This identification was carried out taking source from the work of Bolognesi \cite{bolognesi2013estimer}. Four sets of criteria are then identified: snow and meteorological conditions, the characteristics of the terrain, the characteristics of the practitioner, and the characteristics of the group. The perception of these criteria by off-piste skiers was measured. In addition, it has been observed that the use of different types of criteria varies depending on the availability of avalanche accidents.

We wish to test whether the perceived information, coming from the environment, is heuristically distorted; based on \cite{gletty2017comprendre}. We postulate the following hypothesis:
\begin{quote}
\textbf{Hypothesis 2:} The use of environmental criteria in the decision to go off-piste varies depending on the availability of an avalanche accident event.
\end{quote}

\subsubsection{Heuristic traps in off-piste skiing}

The availability heuristic is well documented in the decision making and risk perception literature. However, there is an approach to heuristics more specific to off-piste skiing. McCammon introduces the notion of heuristic traps among off-piste practitioners \cite{mccammon2004heuristic}. From a manual analysis of 715 avalanche accident reports, he interprets six heuristic traps: 
\begin{itemize}
    \item familiarity, corresponds to the willingness to engage more on a known slope despite an avalanche risk 
    \item the search for rarity, willingness to enter a slope where there is no trace
    \item consistency, willingness to enter a slope given an intention to travel
    \item social facilitation, available to engage in a slope because other skiers are present there
    \item the expert halo, ready to trust a leader according to qualities that differ from his knowledge of avalanche risk
    \item social acceptance, the willingness to go down a slope because it is socially valued.
\end{itemize}
The social dimension in off-piste decision making is largely emphasized here. Moreover, the heuristic of familiarity can be related to the heuristic of availability. Indeed it refers to the availability of knowledge relating to the descent situation.

We want to test the effect of the heuristic of familiarity by integrating it into the social dimension of off-piste skiing. So we postulate:
\begin{quote}
\textbf{Hypothesis 3:} The presence of an individual in a group familiar with an avalanche risk route increases his probability of embarking on this route.
\end{quote}

\section{Methodology}

In order to introduce the methodological approach, we would like to point out that the design of the design and the experimental material was done jointly. However, for the sake of clarity, we break down these two aspects disjointly.

For information, the experimental tool is a research object in its own right. It is a software developed specifically for the experiment, as a \textit{point and click} game, accessible on the web\footnote{The game can be played online (in French) at: \url{http://cartodialect.imag.fr/avalanches-scco/}}. The software simulates an off-piste ski descent through an interactive scenario.

\subsection{Participants and ethics}
The recruitment of participants was done from private messages, scientific mailing lists and as well as through an advertisement on a specialized ski forum \footnote{\url{http://www.skitour.fr/}}. The population we are studying are off-piste practitioners. Different criteria were chosen to select the participants: the level of skiing, the nature of skiing practice (leisure, competition, professional, rescue), the level of off-piste skiing and the frequency of off-piste skiing.

Based on the selection criteria, among the 334 entries, 278 are deemed valid. In the selected sample, we observe that 96\% of participants judge to have a good or excellent level in skiing, 88\% consider to have a good or excellent level in off-piste skiing, 89\% of participants practice off-piste often or very often, and 91\% go off-piste only for leisure. The 4 age groups identified (under 30 years old; 30-40 years old; 40-50 years old; over 50 years old) are distributed fairly evenly, \ie around 25\% (+ or - 2) in each category. The gender of the participants is 87\% male.

At the start of the experiment, the participants validated a consent form, following the recommendations of the CERGA (Research Ethics Committee of the University of Grenoble-Alpes). The data is anonymous, it is hosted in France by Grenoble Informatics Laboratory, and only accessible by project researchers.

\subsection{Experimental design}

\subsubsection{Experimental manipulation of independent variables}

\paragraph{Availability heuristic}

The experiment aims to test the effect of the availability of an avalanche accident event on the probability of engaging in an avalanche risk route (Hypothesis 1). We estimate that in a simulated scenario d In an off-piste ski descent situation, participants viewing an avalanche video choose to forgo downhill skiing more frequently than those viewing a ski descent video without avalanche. In short, we are using video visualization to operationalize the avalanche accident event. 

Thus 3 modalities are posed allowing to study the availability of an avalanche accident event: 
\begin{itemize}
    \item A video containing an avalanche accident \footnote{\url{https://youtu.be/YW5AcukbD3k}}. It consists of a first-person off-piste skiing downhill scene, followed by an avalanche onset and then burial. In a second step, this scene in first person view is put in parallel with the view of the teammate. Although this is not present in the excerpt that we broadcast, the injured person was rescued following the event and escaped unscathed; this is specified at the end of the experiment. 
    \item A video containing ski descents without avalanche \footnote{\url{https://youtu.be/4sY0W9s7puM}}. This video is a compilation of ski runs. For the sake of standardization, the two video clips are of equal length, exactly 2 minutes, they come from the same production collective (ODOS project), and the sound is muted.
    \item The absence of video, which corresponds to a neutral situation.
\end{itemize}

\paragraph{Use of environmental criteria}

Second, the experiment aims to test the effect of the availability of an avalanche accident event on the use of environmental criteria (Hypothesis 2). Thus, different types of criteria relating to off-piste decision making, identified by \cite{gletty2017comprendre} are integrated into the simulation. The data used in modelling the criteria comes from the work of Gletty, based on the responses to a questionnaire aimed at assessing whether the different types of criteria are taken into account in the decision to practice off-piste or to give up. We use the mean response scores on a 5-point Likert-type scale (1: never; 2: rarely; 3: often; 4: very often; 5: always).

\subparagraph{Criteria relating to the characteristics of the land}

This type of criterion is represented by a score of 3.35 out of 5. Two sub-criteria predominate: the presence of elements specific to the terrain (trees, rocky bars, crevasses, lakes, rivers, etc.) with a score of 3.73 out of 5, and the presence of return areas to the ski area with a score of 3.72 out of 5. We set the first criterion by indicating the absence of trees, rocky bars and crevasses. We set the second criterion by indicating a possible return to the ski station at the start and at the end of the route, and the impossibility of returning at the middle of the route. This information is one element that links the scenario to the experience. The decision-making measure takes place when the participant's engagement in the slope can hinder his return to the ski station.

\subparagraph{Criteria relating to the practitioner and the group}
The criteria relating to the practitioner are represented by a score of 3.57 out of 5. We eliminate them from the simulation because they are specific to the participant. The randomization of the experimental groups allows the random distribution of the different profiles of participants.
The criteria relating to the group are represented by a score of 3.35 out of 5. They are the subject of a particular formalization making it possible to test hypotheses 2 and 3.

\subparagraph{Operational hypothesis}
Regarding the variations in the use of environmental criteria in the decision to practice off-piste depending on the availability of an accident event (Hypothesis 2), it has been shown that skiers having already been victims of an avalanche no longer use the criteria related to the group. Indeed, the measures of Gletty are significant in this regard. Quantitatively this translates as follows: skiers without avalanche experience judge to use the criteria of the group with a score of 3.16 out of 5, while in skiers victims of an avalanche it is measured with a score of 3.66 out of 5.

As previously described, we reproduce the ski accident experience by viewing a video. We reduce the use of environmental criteria to the use of a criterion relating to the group: the level of knowledge of the place associated with the availability heuristic that we describe in the next section. Consequently, hypothesis 2 of our experience is operationalized by the effect of interaction between the type of video viewed and the level of knowledge of the place by the group.

\subsubsection{Group characteristic and familiarity heuristic}

Third, the experiment tests the effect of the presence of an individual in a group familiar with an avalanche risk route, on the probability of embarking on that route (Hypothesis 3). The group is represented within the visual scenes by the integration of four characters. Each character is associated with a set of features, formalized by progress bars from 0 to 100.

The importance of the technical level of the members of the group is evaluated by a score of 4.12 out of 5, the importance of the off-piste experience by a score of 3.84 out of 5, the importance of physical condition by a score of 3.22 out of 5. These criteria are randomized by an average level of 50 out of 100, formalized by the function 30 + Z, where Z is a random function, defined on a support bounded between 0 and 40. Although the nature of the representation under form of level is not ecological, we believe that the introduction of variations, intra- and inter-members of the group makes the representation more natural.

The fourth criterion is the knowledge of the place by the group, it refers to the heuristic of familiarity. This variable is made up of two modalities: a weak knowledge of the place, fixed at a level of 10 out of 100 for all the members of the group, and a strong knowledge of the place, fixed at a level of 85 out of 100 for all the members of the group. We are therefore trying to test whether, in a simulated off-piste skiing situation, information on the group strong knowledge of the place promotes the participant's probability of engagement.

\subsection{Measures}

\paragraph{Likelihood of engagement in the situation}
As formalized in the theoretical approach, we use as an indicator of the effect of heuristics, the probability of engaging in an avalanche risk route. In the assessment task, the participant must enter this probability on a bar from 0 to 100 using a slider.

\paragraph{Binary choices}
We have introduced into the experiment another type of evaluation, not associated with hypotheses, which will be the subject of exploratory work. Thus another measure of decision making is carried out through three binary choice tasks.

\begin{enumerate}
	\item Following the assessment of the probability of entering the slope, the participant has the choice of \textit{giving up} or \textit{not giving up} to continue. If the participant enters the slope, he continues the simulation, otherwise it ends.
	\item If he does not give up, a new scene is displayed associated with a choice of trajectory. A dilemma arises where on the left the snow is less good and where on the right the descent is made upstream of the other skiers, behaviour associated with danger.
	\item A last choice is then proposed, it relates to the engagement in a ski jump, with or without cooperation.
\end{enumerate}

\subsection{Procedure}
Through the interface of the experimental software, participants operate in a simulated ski descent environment that integrates static visual scenes. A text associated with each scene provides script information and experimental instructions. In addition, additional components, which can integrate specific information, or make it possible to perform a measurement through an interaction, can be associated with a scene. The experiment can only be performed on a computer with a resolution greater than 1366 * 768 pixels. In order to standardize the display, the graphical interface is \textit{responsive}, that is to say that the display proportions of the elements are the same regardless of the resolution. In order to control the unique participation in the experiment (and yet allow people to replay if wanted), a system of declaration of new participation has been put in place.

There are six experimental conditions, following a mixed design with three contexts (video with avalanche vs. video without avalanche vs. without video) and two conditions (weak familiarity vs. strong familiarity). Each participant is randomly assigned one of the six experimental conditions. The experience takes place through a scenario. At the start of this, the participant is on a route preceding the off-piste, and then goes along the following stages:
\begin{enumerate}
    \item becoming aware of the snow and weather conditions as well as the characteristics of the terrain;
    \item being assigned to one of three contexts;
    \item learning about the characteristics of the group members;
    \item performing a task of evaluating the probability of entering a slope, before being faced with a binary choice (return to the track vs. off-piste engagement) where he is asked to carry out the probabilistic assessment; he then has the opportunity to be reminded about the characteristics of the group members;
    \item making one or more binary choices concerning the descent, that can lead to the onset of an avalanche;
    \item receiving a \textit{feedback}, at the end of the experience, about choices made as well as information on off-piste skiing heuristics.
\end{enumerate}

\subsection{Experimental material}

The design of the experimental equipment is a characteristic point of this study. The objective is to test different hypotheses on decision-making in off-piste skiing in order to establish a model of the risk-taking behaviour of victims of mountain accidents. Moreover, since mountain accidents are undesirable events, a field study is difficult. Therefore, we have chosen to develop a simulated environment, by computer. In order to make it accessible and to stimulate participation in the experiment, we have chosen a support in the form of a game. To do this, we have created an experimental software integrating game mechanisms and dedicated to testing the hypotheses postulated here.

\subsubsection{Serious Game}

\paragraph{Definitions}

First of all, it is necessary to define what the serious game represents. In his work, \cite{alvarez2007jeu} lists two definitions.
The first definition is proposed by \cite{zyda2005visual}, director of the GamePipe laboratory, who sees the \textit{serious game} as “A brain challenge, played with a computer according to specific rules, which uses entertainment as added value for training and training in institutional or private settings, in the fields of education, health, civil security, as well as for communication strategy purposes.”

The second definition, proposed by \cite{sawyer2007}, co-director of the Serious Games Initiative, considers that these are computer applications, produced by "developers, researchers, manufacturers, who look at how to use video games and associated technologies other than entertainment". This is about the terminology of \textit{serious game}, however a nuance can be made by introducing the notion of \textit{gamification} which refers to the use of video game mechanisms within an application, without it necessarily being considered as a video game.

We have reduced our approach to the field of serious games within cognitive sciences. In this context, serious game are considered useful for various objectives. They makes it possible to engage a participant in a task, to collect data on their behaviour, or even to encourage behaviour change \cite{lumsden2016gamification}.

As stated, one of the goals of \textit{gamification} is to increase the motivation and commitment of a participant when performing an experimental task. Indeed, the majority of studies involving behavioural assessment tasks and \textit{gamified} elements have been designed in this objective \cite{lumsden2016gamification}. Thus the aim is to promote participant engagement, to encourage personal motivation \cite{ryan2006motivational}, and to reduce task anxiety. It can be noted that unlike a classical video game, an experimental task is not intended to be replayed outside the associated study.

Two uses of video games are identified in cognitive sciences: cognitive evaluation, and cognitive training \cite{lumsden2016gamification}. Cognitive assessment is an experiment that aims to measure behaviour. Although limited in the changes made to the participant, it allows behavioural measurements to be carried out and requires less resources than the second use. One of the validation standards is then the comparison of a non-game task with a playable task \cite {lumsden2016gamification}. The major difference identified in Lumsden's work between these two types of game is the quality of the final rendering. A training game is intended to be used repetitively, because a cognitive skill is worked on in successive sessions: games in this category therefore usually have an advanced appearance and features. In contrast, the games intended for evaluation are simpler, with the user participating in only one session.

MACBETH\footnote{Mitigating Analyst Cognitive Bias by Eliminating Task Heuristics}, developed by \cite{dunbar2014implicit} is an example of a video game that induces changes in behaviour, both observed and measured. The game tries to mitigate cognitive biases: confirmation bias and fundamental attribution error. Confirmation bias refers to the likelihood of relying more on an initial hypothesis in reasoning, regardless of its truthfulness. The fundamental attribution error is the tendency to place more weight on an individual's internal characteristics rather than environmental factors in the analysis of behaviour. These are therefore two cognitive processes resulting from heuristic processing which are evaluated and whose effects MACBETH aims to attenuate. The game incorporates a scenario and mini-games through which the participant learns to avoid bias. Within the game's mechanics are implemented hypothesis testing and associated measurements.

Finally, we base ourselves on these different theoretical supports to design the experimental material allowing to test our hypotheses. We therefore consider that participant engagement is a major issue in the dissemination of an unpaid online experience. In addition, we believe that the development of video games is an adequate support for the development of a simulation.

\paragraph{Design}
Our study consists of testing the effect of heuristic processing on the choice of a route. It is then a question of performing behavioural measurements and we therefore choose as standard the use of the game on a cognitive evaluation task.

As already specified, one of the validation standards is the comparison of a non-game task with a playable task. However, the specificity of the problem we are studying did not allow us to identify \textit{gamifiable} cognitive evaluation tasks. On the other hand, in order to get closer to the established validation standard, we sought to reproduce observations made in non-game. This is why hypotheses 1 and 2 that we are testing relate to a model experiment which assesses the taking into account of the different types of criteria in the decision to practice off-piste, through the questionnaire defined by \cite{gletty2017comprendre}.

Some elements of the model experiment were then \textit{gamified}. While the avalanche accident experience had been tested as an invoked variable, we propose through the game to transform it into a provoked variable. This means that we change the experience of an accident experienced in a virtual accident. The design of our \textit{gamified} experiment requires a simulated environment, which required us to model and simplify the real environment. In this case, therefore, it is the model of an off-piste skiing situation that we have designed. It is from the experimental material of the model experiment that we designed the game environment. Thus we integrated the snow-meteorological conditions, the characteristics of the terrain, and the characteristics specific to the group.

Furthermore, we do not use the same metrics as the model experiment, so the statistical results of the two experiments cannot be compared and only qualitative interpretations can then provide a possible parallel. We do not cover the creative aspects of the design here because they are not part of a scientific methodology. Ultimately the quantitative evaluation of \textit{gamification} is not possible in our study.

While our study focuses only on behavioural measures, we believe that training human cognition about heuristics would provide a comprehensive prevention game. Our study is therefore positioned as a first step towards the creation of a game that trains cognition with a preventive goal.

\subsubsection{Implementation}
By combining the constraints of experimental design with the game-based approach, we produced dedicated software. It allows experimentation on decision-making in off-piste skiing situations and integrates \textit{gamified} mechanisms. The development required technical choices adapted to the problem as well as the integration of experimental variables into the user interface. The source code is open and available online\footnote{url{https://github.com/Nojann/avalanche-scco.git}}.

\paragraph{Technical details}
In order to provide the participant with an interactive and accessible experience, we used a client-server architecture, implemented according to World Wide Web standards. JavaScript \cite{hazaelmassieux} allows script execution by a web browser and it makes the HTML markup language \cite{htmlstandard_2020} dynamic. In addition to specific languages, our architecture is made up of three services:
\begin{itemize}
    \item Node.js, a server-side JavaScript environment managing user connections \cite{tilkov2010node}.
    \item MongoDB, a document-oriented database storing the application data \cite{chodorow2013mongodb}.
    \item A client-side service, built according to a model-view-controller architecture pattern, intended for the management of the graphical interface, designed with the Angular framework \cite{Jain2014AngularJSAM}.
\end{itemize}

In addition, the data format used is JavaScript Object Notation (JSON \cite {bray2014javascript}). It is used across the entire information processing chain: client-side data capture, server-side storage, and statistical processing with R\footnote{\url{https://www.r-project.org/about.html}}.

The implementation has mainly focused on client-side service. Indeed, it is through a user interface that the experience unfolds. A set of Web components \cite{Glazkov:14:IWC} was then designed, allowing display and interaction. The design of the components was carried out using the Bootstrap framework \cite{spurlock2013bootstrap} and the Angular Material component library\footnote{\url{https://material.angular.io/}}.

\paragraph{Experimental software}
The software is divided into two parts: the game engine and the scene editor.
The game engine calculates and organizes the display of the various components through a grid. The game is thus composed of 20 grids, systematically comprising an illustrated scene and a dialog box. The latter displays textual script elements as well as the instructions for the experiment. Various optional components can be associated with the grid. They may have the role of making information accessible or retrieving it by interacting with the participant. On the one hand, several components display information: a component implements the environmental criteria (snow-meteorological conditions and characteristics of the terrain), a component implements the characteristics of each member of the group, and a component provides a visual scene integrating a background and different characters. On the other hand, a set of components is dedicated to the interaction: a component allows the entry of the participant's personal data and a component implements the evaluation of the probability of engagement, a component implements the choice between two options. Some components are associated with a dedicated calculation engine allowing complex information processing.

\begin{figure}[hbt]
\begin{center}
    \includegraphics[width=140mm]{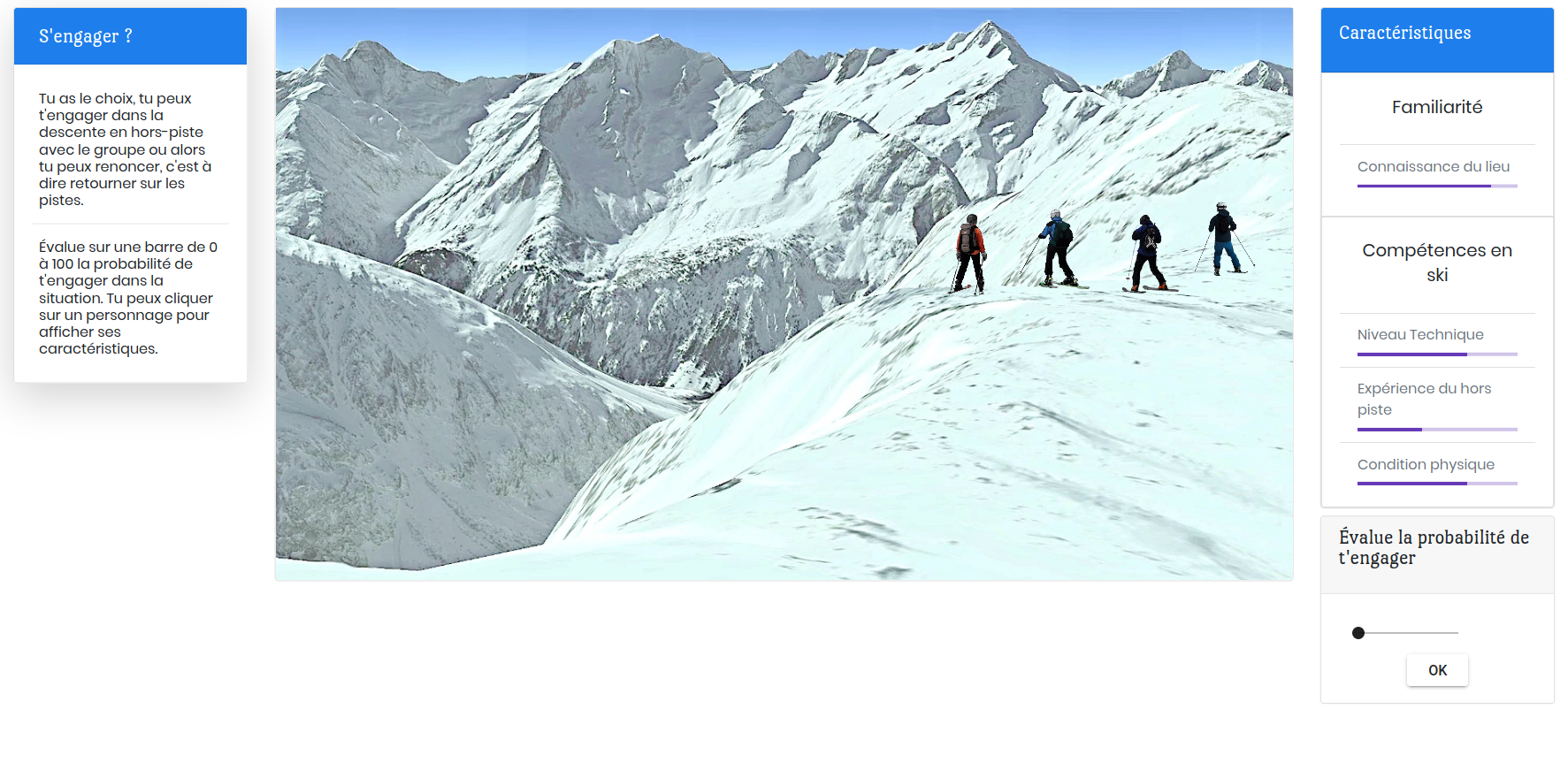}
    \caption{Example components grid} \label{fig:screen}
\end{center}
\end{figure}

\newpage
\noindent Figure~\ref{fig:screen} is a screenshot of our experimental software\footnote{\url{http://cartodialect.imag.fr/avalanches-scco/}}. It shows the following components:
    \begin{itemize}
    		\item Left component: textual scenaristic elements, instructions
    		\item Central component: illustrated scene with 4 characters
    		\item Upper right component: features of the selected group member
    		\item Lower right component: engagement probability measure
    \end{itemize}
    
\section{Results}

\subsection{Sample}
We recorded 334 total observations. Among these, we removed 43 observations representing a repeat of the experiment, and 13 valid observations but corresponding to a participant with a low level of skiing or not practicing off-piste. The retained sample has a size of 278, distributed across the different conditions as shown in Table~\ref{tab:sample} below.

\begin{table}[hb]
    \centering
    \begin{tabular}{|c|c|c|c|}
        \hline
        Condition & Avalanche video & No avalanche video & No video \\
        \hline
        Low Familiarity & 46 & 40 & 43 \\
        \hline
        High Familiarity & 56 & 44 & 49 \\
        \hline
    \end{tabular}
    \caption{Distribution of the sample over the 6 conditions}
    \label{tab:sample}
\end{table}

Regarding the probability of engagement of the whole sample, with a measurement range of 0 to 100, the mean is 50.23, the median is 50, the standard deviation is 25.65.

\subsection{ANOVA validity hypotheses}
The ANOVA validity assumptions were tested. There is no repetition of measurements in the experimental design and the independence of the residuals is validated graphically. A Shapiro-Wilk test is performed ($p = .06$). The null hypothesis, which states that the population is normally distributed, is not rejected ($\alpha = .05$) and we conclude that the residuals are normal. A Levene test is performed ($p = .11$). The null hypothesis, which states that the variances of the different groups are broadly identical, is not rejected ($\alpha = .05$) and we conclude that the variances are homogeneous.

\subsection{Interaction}
We perform a type III ANOVA on the probability of engagement given three contexts (avalanche vs. no avalanche vs. no video) and two conditions (low familiarity vs. high familiarity). This results in an absence of interaction ($p = .29$) (see Table~\ref{tab:anova}).

\begin{table}[ht]
\centering
\begin{tabular}{lrrrr}
  \hline
 & Sum Sq & Df & F value & Pr($>$F) \\ 
  \hline
  (Intercept) & 696641.97 & 1 & 1102.28 & 0.0000 \\ 
  videoType & 5125.79 & 2 & 4.06 & 0.0184 \\ 
  familiarity & 3723.70 & 1 & 5.89 & 0.0159 \\ 
  videoType:familiarity & 1554.00 & 2 & 1.23 & 0.2941 \\ 
  Residuals & 171904.45 & 272 &  &  \\ 
   \hline
\end{tabular}
\caption{ANOVA results, generated with R} \label{tab:anova}
\end{table}

Given the non-significance of the interaction, we fit the model, without testing the interaction, by performing a type II ANOVA.

\subsection{Main effects}
From an ANOVA, we observe a main effect of the context, with F(2,275) = 4.28, and $p$ = .01. The situational context includes three modalities: with avalanche video (M = 45.2, SD = 22.46), with video without avalanche (M = 55.82, SD = 25.73), without video (M = 50.72, SD = 27.97).

We observe a main effect of the group's familiarity with the location, F(1,276) = 5.66, and $p = .02$. Group familiarity includes 2 modalities: group with strong knowledge of the place (M = 53.49, SD = 25.07), group with low knowledge of the place (M = 46.47, SD = 25.90).

\subsection{Comparison of the means}
In order to perform a two-by-two comparison of the means, we use a Tukey range test. The results are shown in Table~\ref{tab:tukey}. The comparisons of the Avalanche / No video and No avalanche / No video pairs do not show any significant differences ($p > .05$). 

\begin{table}[ht]
\centering
\begin{tabular}{rrrrr}
  \hline
 	& Est. Std. & Error & t value & Pr($>|t|$) \\ 
  \hline
  Avalanche - No avalanche  & 10.81 & 3.71 & 2.91 & 0.01 \\ 
  Avalanche - No video & 5.64 & 3.62 & 1.56 & 0.35 \\ 
  No avalanche - No video & -5.17 & 3.80 & -1.36 & 0.47 \\ 
  Strong / Low familiarity & 7.20 & 3.03 & 2.38 & 0.06 \\ 
  \hline
  \end{tabular}
  \caption{Tukey test results} \label{tab:tukey}
\end{table}

\begin{figure}[hbt]
\begin{center}
    \includegraphics[width=120mm]{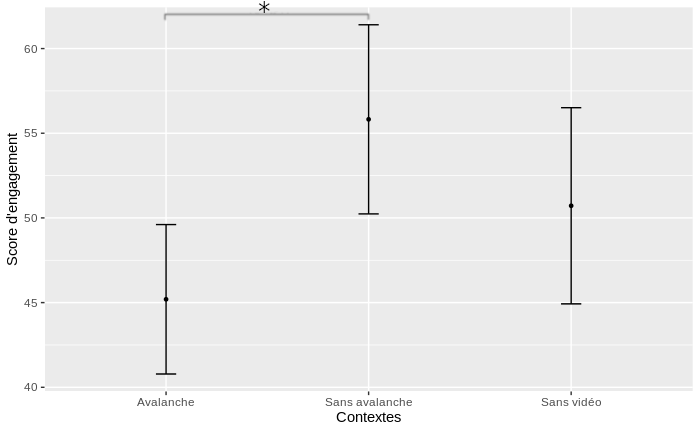}
    \caption{Slope engagement probability depending on context} \label{fig:prob}
\end{center}
\end{figure}

On the other hand, we observe (see Figure~\ref{fig:prob}) that the probability of engagement in a context without avalanche (M = 55.82, SD = 25.73) is significantly higher ($p  = .01$) than the probability of engagement in a context with avalanche (M = 45.2, SD = 22.46).

A priori, the comparison of the Strong Familiarity / Weak Familiarity pair does not show any significant differences ($p > .05$). However, with a risk of error ($\alpha$) of 0.1, we observe that the probability of engagement given a strong familiarity (M = 53.49, SD = 25.07) is significantly higher ($p = .06$) than the probability of engagement given low familiarity (M = 46.47, SD = 25.90). Given the risk of errors at 10\% rejection of the null hypothesis ($\mu$ \textit{strong familiarity} = $\mu$ \textit{weak familiarity}), this result should be interpreted with caution.

\section{Discussion}
\subsection{Behaviour of mountain accidents victims}

\subsubsection{Effect of the availability heuristic}
We tested the effect of the availability of an avalanche accident event on the probability of engaging in a route (Hypothesis 1). This effect is associated with the availability heuristic \cite{tversky1973availability} and has been observed in an off-piste skiing situation \cite{gletty2017comprendre}. We introduced the availability of the avalanche accident event through three different situational contexts: avalanche video, avalanche free video, and no video. We measure the effect on the probability of engaging in an avalanche risk route associated with the perception of risk.

As a result of the experiment, we did not observe any significant differences between viewing different videos and no video. Within our simulation, the availability of the simulated event through video therefore has a negligible effect. The comparison withGletty's observations suggests that the operationalization of an accident event by a video does not have an effect similar to a lived situation. We consider this to be a limitation of our simulation, realized by video support, with respect to a real situation. Indeed, the practice of skiing mobilizes a set of sensory sensors and motor reactions, while viewing a (silent) video can only be considered to mobilize visual perception. While it is undesirable to test the effect of an avalanche accident in a real-life situation, it is possible that a simulated environment, more comprehensive than that developed here, will increase the effect of availability. The use of virtual reality headsets as well as the introduction of feedbacks on all sensory channels is an avenue to explore to increase the ecology of the task and the associated effects \cite{cipresso2018past}.

Although the effect of a video versus no video was not seen, we observed a significant difference between the two simulated events across the videos. In fact, viewing video containing an avalanche accident is associated with a lower probability of engagement than video containing ski descents without an avalanche. In this context, we conclude that the availability of the avalanche accident event favors the probability of giving up on an off-piste route. Hypothesis 1 is therefore not rejected and the observations are in accordance with previous works by \cite{tversky1973availability} on the availability heuristic, as well as \cite{gletty2017comprendre} on the availability heuristic for off-piste skiing.

\subsubsection{Use of environmental criteria}

We tested the effect of availability on the use of environmental criteria in the decision to go off-piste (hypothesis 2). This effect is in line with Gletty's observations showing a significant difference in the evaluation of characteristics relating to the group according to the avalanche experience of the practitioners. 

We then tested the presence of an interaction effect between the type of video viewed and the group's level of knowledge of the place. We have not observed here any interaction between the situational context and the familiarity criterion. We therefore reject hypothesis 2. In Gletty's experiment, risk perception measurements were carried out using a questionnaire in which the participant is asked to evaluate different off-piste decision criteria. In contrast, we assessed the perception of risk implicitly, through the interaction of two independent variables. By introducing this test into a simulated environment and performing an implicit measurement for the participant, we do not observe the previously identified effects. We attribute this disparity to a different use of the criteria in the experimental method. Indeed, in Gletty's experiment, risk perception is measured by a direct evaluation of the criteria by the participant, while in the experiment that we propose, the criteria are represented and quantified and we measure the effect of their variation. We then conclude that further action is needed to determine how off-piste decision criteria are perceived, given the availability of an avalanche accident.

\subsubsection{Familiarity heuristic}

We tested the effect of the presence of a participant in a group familiar with the location of the activity, postulating that familiarity promotes the probability of embarking on a risky route (hypothesis 3). This effect is associated with the familiarity heuristic \cite{mccammon2004heuristic} that we operationalized in two modalities: strong knowledge of the place by the group vs. little knowledge of the place. We observe a significant effect of familiarity which has an effect on the probability of engagement. The likelihood of engagement tends to be higher when the place is familiar than when it is not. Therefore, we do not reject hypothesis 3.

While we did not observe any effect of the avalanche accident event on the perception of the criteria relating to the group, we observe that the criteria relating to the group are influencing in the decision of s 'enter a route. This observation is in line with the work of McCammon and Gletty. On the one hand, according to McCammon, familiarity corresponds to the willingness to go further down a known slope despite the risk of an avalanche. On the other hand, Gletty associates familiarity with the group of skiers. We conclude that the heuristic familiarity trap, present during an off-piste ski descent, has been reproduced by the simulation we designed.

\subsubsection{Behaviour model}

Given that we observed the effect of availability and familiarity heuristics on the probability of engaging in an avalanche risk situation, we conclude on the effect of heuristic processing in off-piste skiing. The choice of an itinerary operated by skiers is subject to identifiable errors in judgment. We believe that it is possible to infer the off-piste ski route choice situation from route choices made in other mountain activities. We therefore consider that risk-taking leading to accidents in the mountains is a consequence of the distortion operated by heuristic processing on the perception of the frequency of occurrence of risky events.

\subsection{Experimental software}

The design of the experimental material through a \textit{gamified} software was intended to stimulate participation in the experiment, provide a simulated environment, and design the first stage of a cognition-training game.

Having no metric to measure the impact of participant engagement in the game, we offer a subjective analysis. The software was deployed online over a three-week period and we obtained 334 entries of which 278 were valid, with an average of 46 participants per condition. As the study took place at the end of the ski season, with special sanitary conditions (gradual lifting of sanitary lockdown measures), we believe we had a satisfactory number of participants, ensuring the validity of our results. The dissemination process took place through a virtuous circle of sharing within communities of skiers. 

We observed an effect of the availability heuristic and the familiarity heuristic on the decision-making to practice an off-piste route. These observations are in agreement with the experiment used as a model \cite{gletty2017comprendre}. However, we do not observe a similar use of decision criteria. We believe that we have enough observations of identical effects to conclude that the \textit{gamification} process has been successful, which is in line with \cite{lumsden2016gamification}. Various qualitative feedback from skiing and mountain rescuing professionals also confirm that the interactive form of the experience is an asset.

Finally, we confirm that the development of this \textit{gamified} experience can be considered as a first step towards a game involving cognition with a preventive goal: the effect of heuristics has been observed. However, it seems necessary to us to increase the quality of the simulation. This can be done by increasing the number of environmental variables and by a more ecological implementation of these variables.

\section{Conclusion}
In order to understand decision-making in mountain activities, we observed the effect of availability and familiarity heuristics on the probability of embarking on a risky route. There was a double challenge here: helping to improve the location of mountain accident victims, and also improving prevention of these accidents.

Our observations led us to consider that risk-taking leading to accidents in the mountains is a consequence of the distortion operated by heuristic processing on the perception of the frequency of occurrence of risky events. In addition, we have shown that the choice of an itinerary is conditioned by judgment heuristics. In this context, the study of cognitive processes can be useful to determine the trajectories of victims of mountain accidents and thus improve their location.

The design of an experimental software \textit{gamified}, allowed us to perform various measurements on heuristic processing. The number of participants and the positive qualitative feedback on this tool open up the prospect of the development of avalanche prevention support in the form of a \textit{serious game}.

\bibliographystyle{apalike}

\begin{thebibliography}{}

\bibitem[Alvarez, 2007]{alvarez2007jeu}
Alvarez, J. (2007).
\newblock {\em Du jeu vid{\'e}o au serious game: approches culturelle,
  pragmatique et formelle}.
\newblock PhD thesis, Toulouse 2.

\bibitem[Bolognesi, 2013]{bolognesi2013estimer}
Bolognesi, R. (2013).
\newblock Estimer et limiter le risque avalanche.

\bibitem[Bray et~al., 2014]{bray2014javascript}
Bray, T. et~al. (2014).
\newblock The javascript object notation (json) data interchange format.
\newblock RFC 7159, DOI 10.17487/RFC7159, March 2014.

\bibitem[Chodorow, 2013]{chodorow2013mongodb}
Chodorow, K. (2013).
\newblock {\em MongoDB: the definitive guide: powerful and scalable data
  storage}.
\newblock " O'Reilly Media, Inc.".

\bibitem[Cipresso et~al., 2018]{cipresso2018past}
Cipresso, P., Giglioli, I. A.~C., Raya, M.~A., and Riva, G. (2018).
\newblock The past, present, and future of virtual and augmented reality
  research: a network and cluster analysis of the literature.
\newblock {\em Frontiers in psychology}, 9:2086.

\bibitem[Dunbar et~al., 2014]{dunbar2014implicit}
Dunbar, N.~E., Miller, C.~H., Adame, B.~J., Elizondo, J., Wilson, S.~N., Lane,
  B.~L., Kauffman, A.~A., Bessarabova, E., Jensen, M.~L., Straub, S.~K., et~al.
  (2014).
\newblock Implicit and explicit training in the mitigation of cognitive bias
  through the use of a serious game.
\newblock {\em Computers in Human Behavior}, 37:307--318.

\bibitem[Evans, 2008]{evans2008dual}
Evans, J. S.~B. (2008).
\newblock Dual-processing accounts of reasoning, judgment, and social
  cognition.
\newblock {\em Annu. Rev. Psychol.}, 59:255--278.

\bibitem[Glazkov and Ito, 2014]{Glazkov:14:IWC}
Glazkov, D. and Ito, H. (2014).
\newblock Introduction to web components.
\newblock {WD} not longer in development, W3C.
\newblock http://www.w3.org/TR/2014/NOTE-components-intro-20140724/.

\bibitem[Gletty, 2017]{gletty2017comprendre}
Gletty, M. (2017).
\newblock {\em Comprendre les pratiques, les perceptions, l’explication
  na{\"\i}ve des accidents, les croyances relatives au risque d’avalanche
  pour mieux pr{\'e}venir les accidents en hors-piste chez les jeunes
  pratiquants de sports de glisse}.
\newblock PhD thesis, University Grenoble Alpes.

\bibitem[Group, 2020]{htmlstandard_2020}
Group, W. H. A. T.~W. (2020).
\newblock Html living standard.

\bibitem[Hazael-Massieux, 2020]{hazaelmassieux}
Hazael-Massieux (Accessed 15 June 2020).
\newblock Javascript web apis.

\bibitem[Jain et~al., 2014]{Jain2014AngularJSAM}
Jain, N., Mangal, P., and Mehta, D. (2014).
\newblock Angularjs: A modern mvc framework in javascript.
\newblock {\em Journal of Global Research in Computer Sciences}, 5:17--23.

\bibitem[Kahneman and Frederick, 2002]{kahneman2002representativeness}
Kahneman, D. and Frederick, S. (2002).
\newblock Representativeness revisited: Attribute substitution in intuitive
  judgment.
\newblock {\em Heuristics and biases: The psychology of intuitive judgment},
  49:81.

\bibitem[Leplat, 2006]{leplat2006risque}
Leplat, J. (2006).
\newblock Risque et perception du risque dans l’activit{\'e}.
\newblock {\em Psychologie du risque: identifier, {\'e}valuer, pr{\'e}venir},
  pages 19--33.

\bibitem[Lumsden et~al., 2016]{lumsden2016gamification}
Lumsden, J., Edwards, E.~A., Lawrence, N.~S., Coyle, D., and Munaf{\`o}, M.~R.
  (2016).
\newblock Gamification of cognitive assessment and cognitive training: a
  systematic review of applications and efficacy.
\newblock {\em JMIR serious games}, 4(2):e11.

\bibitem[McCammon, 2004]{mccammon2004heuristic}
McCammon, I. (2004).
\newblock Heuristic traps in recreational avalanche accidents: Evidence and
  implications.
\newblock {\em Avalanche news}, 68(1):42--50.

\bibitem[Meunier, 2016]{meunier2016raisonnement}
Meunier, J.-M. (2016).
\newblock {\em Raisonnement, r{\'e}solution de probl{\`e}mes et prise de
  d{\'e}cision}.
\newblock Dunod.

\bibitem[Pachur et~al., 2012]{pachur2012people}
Pachur, T., Hertwig, R., and Steinmann, F. (2012).
\newblock How do people judge risks: availability heuristic, affect heuristic,
  or both?
\newblock {\em Journal of Experimental Psychology: Applied}, 18(3):314.

\bibitem[Ryan et~al., 2006]{ryan2006motivational}
Ryan, R.~M., Rigby, C.~S., and Przybylski, A. (2006).
\newblock The motivational pull of video games: A self-determination theory
  approach.
\newblock {\em Motivation and emotion}, 30(4):344--360.

\bibitem[Sawyer, 2007]{sawyer2007}
Sawyer, B. (2007).
\newblock The "serious games" landscape.
\newblock In {\em Instructional \& Research Technology Symposium for Arts,
  Humanities and Social Sciences}, Camden, USA.

\bibitem[Slovic, 1987]{slovic1987perception}
Slovic, P. (1987).
\newblock Perception of risk.
\newblock {\em Science}, 236(4799):280--285.

\bibitem[Slovic et~al., 2004]{slovic2004risk}
Slovic, P., Finucane, M.~L., Peters, E., and MacGregor, D.~G. (2004).
\newblock Risk as analysis and risk as feelings: Some thoughts about affect,
  reason, risk, and rationality.
\newblock {\em Risk Analysis: An International Journal}, 24(2):311--322.

\bibitem[Slovic, 2000]{slovic2000perception}
Slovic, P.~E. (2000).
\newblock {\em The perception of risk.}
\newblock Earthscan publications.

\bibitem[Soul{\'e} et~al., 2017]{soule2017incidents}
Soul{\'e}, B., Vanpoulle, M., Lef{\`e}vre, B., Boutroy, E., Reynier, V., and
  Routier, G. (2017).
\newblock Incidents et quasi-accidents dans les sports de montagne. premiers
  enseignements et perspectives de pr{\'e}vention.
\newblock Technical report, Petzl Foundation.

\bibitem[Spurlock, 2013]{spurlock2013bootstrap}
Spurlock, J. (2013).
\newblock {\em Bootstrap: Responsive Web Development}.
\newblock " O'Reilly Media, Inc.".

\bibitem[Tilkov and Vinoski, 2010]{tilkov2010node}
Tilkov, S. and Vinoski, S. (2010).
\newblock Node. js: Using javascript to build high-performance network
  programs.
\newblock {\em IEEE Internet Computing}, 14(6):80--83.

\bibitem[Tversky and Kahneman, 1973]{tversky1973availability}
Tversky, A. and Kahneman, D. (1973).
\newblock Availability: A heuristic for judging frequency and probability.
\newblock {\em Cognitive psychology}, 5(2):207--232.

\bibitem[Tversky and Kahneman, 1974]{tversky1974judgment}
Tversky, A. and Kahneman, D. (1974).
\newblock Judgment under uncertainty: Heuristics and biases.
\newblock {\em science}, 185(4157):1124--1131.

\bibitem[Tversky and Kahneman, 1979]{tversky1979prospect}
Tversky, A. and Kahneman, D. (1979).
\newblock Prospect theory: An analysis of decision under risk.
\newblock {\em Econometrica}, 47(2):263--291.

\bibitem[Tversky and Kahneman, 1983]{tversky1983extensional}
Tversky, A. and Kahneman, D. (1983).
\newblock Extensional versus intuitive reasoning: The conjunction fallacy in
  probability judgment.
\newblock {\em Psychological review}, 90(4):293.

\bibitem[Wickens, 1991]{wickens1991processing}
Wickens, C.~D. (1991).
\newblock Processing resources and attention.
\newblock {\em Multiple-task performance}, 1991:3--34.

\bibitem[Zyda, 2005]{zyda2005visual}
Zyda, M. (2005).
\newblock From visual simulation to virtual reality to games.
\newblock {\em Computer}, 38(9):25--32.

\end{thebibliography}

\end{document}